\documentclass{aastex}

\newif\ifEmulate
\usepackage{emulateapj5} \Emulatetrue
\def\Figure#1{\ifEmulate{#1}\fi}

\let\mic=\micron
\def\E#1{\hbox{$10^{#1}$}}
\def\sub#1{_{\rm #1}}
\def\x      {\hbox{$\times$}}
\def\about  {\hbox{$\sim$}}
\def\tV     {\hbox{$\tau\sub V$}}
\def\Rin    {\hbox{$R\sub{in}$}}
\def\Rout   {\hbox{$R\sub{out}$}}

\def\Mdot   {\hbox{$\dot M$}}
\def\Mo     {\hbox{M$_\odot$}}
\def\Lo     {\hbox{L$_\odot$}}
\def\Tc     {\hbox{$T\sub c$}}

\def\Ts     {\hbox{$T_{\ast}$}}
\def\ve     {\hbox{$v\sub e$}}

\def\rgd    {\hbox{$r\sub{gd}$}}
\def\kms    {\hbox{km s$^{-1}$}}

\slugcomment{Revised, September 20, 2000}

  \shorttitle{Water and Dust Emission from W Hydrae}
  \shortauthors{Zubko \& Elitzur}

\begin{document}

\title{WATER AND DUST EMISSION FROM W HYDRAE}

\author{Victor Zubko\altaffilmark{1} and Moshe Elitzur}
\affil{Department of Physics and Astronomy, University of Kentucky,
            Lexington, KY 40506--0055}
  \email{zubko@pa.uky.edu, moshe@pa.uky.edu}

\altaffiltext{1}{On leave from the Main Astronomical Observatory,
           National Academy of Sciences, Kiev, Ukraine}

\begin{abstract}
We construct a self-consistent model for the wind around W Hya by solving the
coupled equations describing the hydrodynamics and dust radiative transfer
problems. The model matches simultaneously the observed continuum radiation and
wind velocity profile. The water line emission is calculated next using the
water abundance as the only free parameter, fitted from the ISO observations of
\citet{neufeld96} and \citet{barlow96}. The gas temperature is determined from
a thermal balance calculation that includes water as one of its main
components. Our model successfully fits all the observed water lines, resolving
a major discrepancy between the modeling results of the two observing teams.
The mass loss rate is 2.3\x\E{-6} \Mo\ yr$^{-1}$, the water abundance is
1.0\x\E{-4} and the ortho:para ratio is 1:1.3.
\end{abstract}


\section{INTRODUCTION}
        \label{sec:introduction}

Water is a dominant coolant of outflows around cool oxygen-rich stars
(Goldreich \& Scoville 1976, GS hereafter; Chen \& Neufeld 1995, CN hereafter;
Truong-Bach et al 1999, TB hereafter). However, until recently observations of
water cooling lines were impossible because of the atmospheric opacity at
these wavelengths. The situation has changed with the successful launch of the
Infrared Space Observatory (ISO). One of the first objects observed with ISO
was W Hydrae, an M7.5 semi-regular red giant, and it showed the expected
thermal water emission in observations with both the SWS (Neufeld et al 1996;
hereafter N96) and LWS instruments (Barlow et al 1996; hereafter B96).  Both
teams also fitted their observations based on the GS approach, resulting in
strikingly different estimates for the mass loss rate: 6$\times$10$^{-7}$
M$_{\odot}$ yr$^{-1}$ (B96) and (0.5--3)$\times$10$^{-5}$ M$_{\odot}$
yr$^{-1}$ (N96). Here we aim to resolve this discrepancy. In contrast with the
original studies we construct a self-consistent model of the radiation field,
dust, and gas in the shell, taking account of all the infrared continuum
observations as an additional constraint.

\section{MODELING}
    \label{sec:modeling}

The driving force of the wind is radiation pressure on the dust, the gas
particles are dragged along by collisions with the dust grains.  The internal
properties of the gas, such as temperature, do not play any role in the
dynamics, the wind structure can be obtained by solving the coupled equations
for hydrodynamics and dust radiative transfer. We now describe our calculation
for W Hya.  With the derived model we proceed to solve the H$_2$O level
population problem.

\subsection{Dynamics and IR Emission}

A complete calculation of the wind structure requires a solution of the coupled
hydrodynamics and dust radiative transfer problems. Traditionally these
calculations involved a large number of input parameters. However,
\citet{ie95} noted that the dusty wind problem possesses general scaling
properties such that, for a given type of grains, both the dynamics and
radiative transfer depend primarily on a single parameter -- the overall
optical depth. Subsequent analysis by \citet{ie97} established rigorously that
the dust radiative transfer problem possesses scaling properties under the
most general circumstances.  Scaling was incorporated in the code
DUSTY\footnote{Accessible at \url{http://www.pa.uky.edu/$\sim$moshe/dusty}}
\citep{dusty}, which solves fully the dusty wind problem. The solution
provides the radial variation of the velocity and radiation fields in terms of
the scaled distance $y = r/\Rin$, where \Rin\ is the shell inner boundary.
That boundary is defined by the condition $\Tc = T_{\rm{dust}}(y = 1)$, where
\Tc\ is the dust condensation temperature. The actual value of \Rin\ never
enters.

We use for modeling ``astronomical silicate'' dust grains with optical
constants from \citet{ld93} and the power law size distribution of \citet{mrn}.
We assume prompt dust formation at $y$ = 1 and no further grain growth or
destruction. The only input parameter in addition to the dust properties is
the stellar temperature $T_\ast$ = 2500~K \citep{hanif95}. From a series of
{\sc DUSTY} models in which we varied the visual optical depth \tV, the
temperature \Tc\ and the shell outer radius $Y$ = \Rout/\Rin\ we chose the one
that best fits all the observations all the way from 1 \mic\ to 1.2 mm; because
of its irregular variability, optical data is not included. We find that
models with  \tV\ = 0.7--1.0, \Tc\ = 900--1100~K and $Y>$ 1000 are almost
equally successful in reproducing the observational spectrum. The most
significant parameter by far is \tV; \Tc\ has only a small effect on the
fitting and the role of $Y$ is marginal. The model with \tV\ = 0.83, \Tc\
=1000~K and $Y$ = 12,000 minimizes the fitting errors and is presented in Fig.
\ref{fig:spectrum}. Matching the model flux to observations in scale as well
as spectral shape determines the shell angular dimensions \citep{ie97}.  The
shell inner diameter is 0\farcs182, consistent with the measured stellar
diameter 0\farcs046 \citep{hanif95}, and its outer diameter 36\arcmin, in
agreement with the IRAS observations of \citet{hawkins90}.

%%%%%%%%%%%%%%%%%%%% Spectrum %%%%%%%%%%%%%%%%%%%%%%%%%%%%%%%%
\Figure{
\centerline{\includegraphics[width=\hsize,clip]{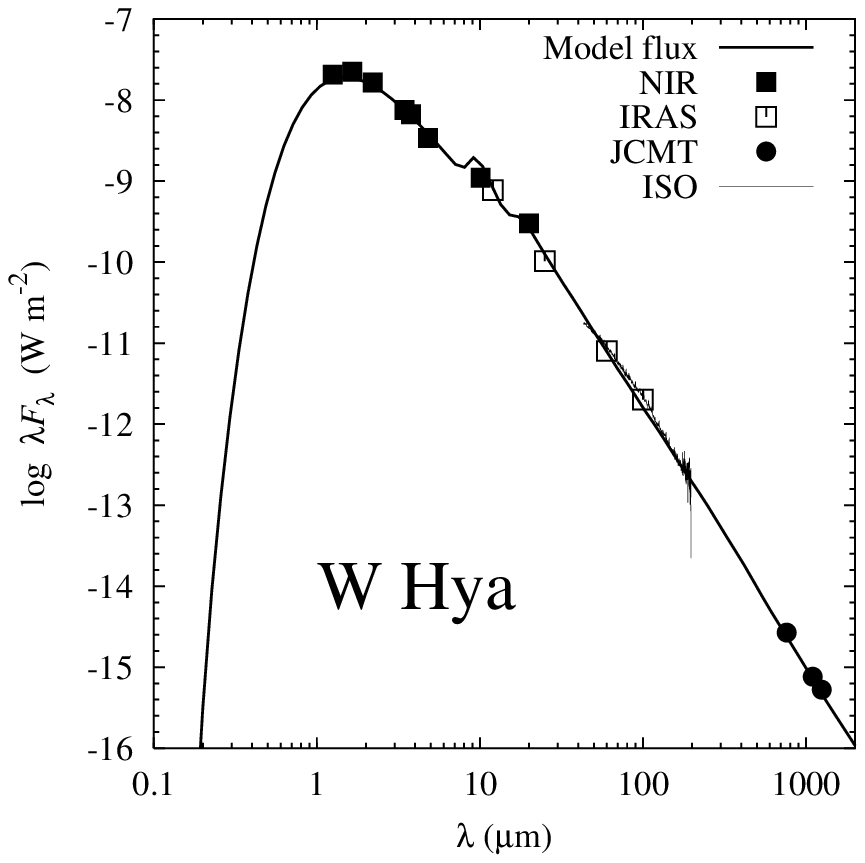}}
\figcaption[spectrum.ps] {The spectral energy distribution of W Hya. Data
indicated by filled squares (NIR) are from \citet{wilson72}; open squares
(IRAS) from \citet{hawkins90}; filled circles (JCMT) from \citet{veen95} and
\citet{walm91}. The thin solid line shows the ISO data of \citet{barlow96}.
The bold solid line is our model result.
      \label{fig:spectrum}}
}
%%%%%%%%%%%%%%%%%%%%%%%%%%%%%%%%%%%%%%%%%%%%%%%%%%%%%%%%%%%%%%

\subsection{Water Lines}
        \label{sec:water_lines}

The statistical rate equations for the water level populations require the
following input: the radiation field, to determine radiative excitations; the
gas density and temperature, for the collision rates; and the water column
density, required for the line radiative transfer \citep{elitzur92}. DUSTY's
output provides the radiation field as well as the dimensionless velocity and
density profiles in the shell. Fixing the scale of densities requires two
additional input properties. We take the distance $d$ = 115 pc from the
Hipparcos catalog. This sets \Rin\ = 1.6\x\E{14} cm and \Rout\ $\sim$ 1 pc, and
the luminosity is 11,050 \Lo\ in agreement with \citet{hanif95}. Next, the wind
terminal velocity \ve\ = 8 \kms\ \citep{young95} fixes the velocity scale and
determines the gas-to-dust mass ratio \rgd\ = 850 and the mass-loss rate \Mdot\
= 2.3\x\E{-6} \Mo\ yr$^{-1}$. The complete velocity profile is shown in 
figure 2
together with the data for OH and H$_2$O masers as well as CO thermal emission.
All are properly explained by the model results. The SiO data is displaced from
the wind velocity profile, as expected for this maser's location inside the
dust formation zone (cf Elitzur 1992).

The only required input quantities that remain unknown are the gas temperature
and water abundance, and we fit those simultaneously from the water line
observations. We calculate the temperature from the balance of cooling and
heating due to adiabatic expansion, grain-gas collisions and H$_2$O
rovibrational transitions (we estimate the H$_2$ vibrational contribution and
find it negligible). We solve for the populations of the lowest 45 rotational
levels of the ground vibrational state of ortho- and para-water; this accounts
for all levels with energy $\la$ 2000 K above ground. The molecular data are
from the HITRAN database \citep{rothman98}, the collision rate coefficients
from \citet{green93}. The level populations and line emissivities are
calculated with the escape probability method as functions of distance $r$,
and the line fluxes by integrating the line emissivities over the shell
volume. The free parameters are the abundances of the two water species, which
must be considered independently since there is no radiative coupling between
them. Detailed modeling shows that these abundances are constant in the region
where the water emission originates (CN, TB). Because of the central role of
H$_2$O in the energy balance, the calculations of the temperature and water
line emission are coupled, and repeated until the best fit achieved for the
line observations. The best fit is found for water abundance $n({\rm
H_2O})/n(\rm H_2)$ = 1.0\x\E{-4} and the ratio ortho:para = 1:1.3. Figure 3
shows the temperature profile. The model parameters are summarized in Table 1,
results and comparison with observations in Table 2.

%%%%%%%%%%%%%%%%%%%% v-profile %%%%%%%%%%%%%%%%%%%%%%%%%%%%%%%%
\Figure{
\bigskip \centerline{\includegraphics[width=\hsize,clip]{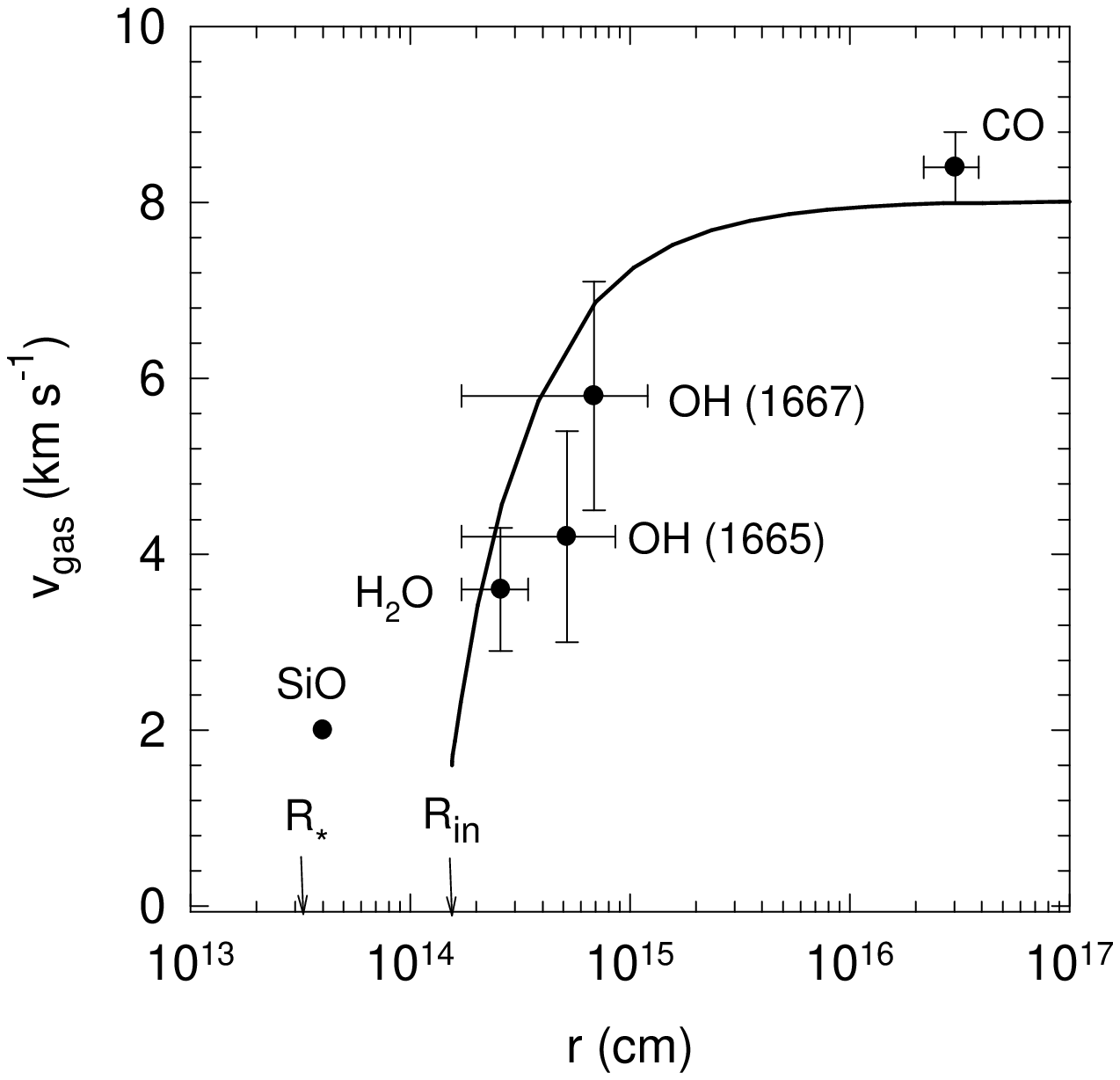}}
\figcaption[vgas.ps] {The model prediction for the gas velocity profile. Data
points (from Szymczak et al, 1998) show the H$_2$O and OH maser and CO thermal
emission from the wind.  The SiO maser emission originates in the extended
atmosphere.
   \label{fig:vgas}}
\medskip
}
%%%%%%%%%%%%%%%%%%%%%%%%%%%%%%%%%%%%%%%%%%%%%%%%%%%%%%%%%%%%%

\section{DISCUSSION}
    \label{sec:discussion}

Our model fits all the water lines within the observational errors, which
generally exceed 50\%. The quality of the fit for the LWS data is comparable to
that of the B96 model. For the SWS data N96 present a range of models for each
line, and our model fits that data set better than any single one of them.
Therefore, our model resolves the conflict among the previous water line
calculations, fitting all the data with a single value for the mass loss rate.
It is important to note that \Mdot\ is determined by the infrared data and \ve,
and remains unchanged during the modeling of the water lines. We find that the
acceptable range of \Mdot\ is \about\ 2--3\x\E{-6} \Mo\ yr$^{-1}$, with a
nominal value of 2.3\x\E{-6} \Mo\ yr$^{-1}$.  Except for the high-end of the
N96 range, most estimates of \Mdot\ are in agreement with ours within the
errors (cf N96 and references therein).

%%%%%%%%%%%%%%%%%%%% T-profiles %%%%%%%%%%%%%%%%%%%%%%%%%%%%%%%%
\Figure{
\centerline{\includegraphics[width=\hsize,clip]{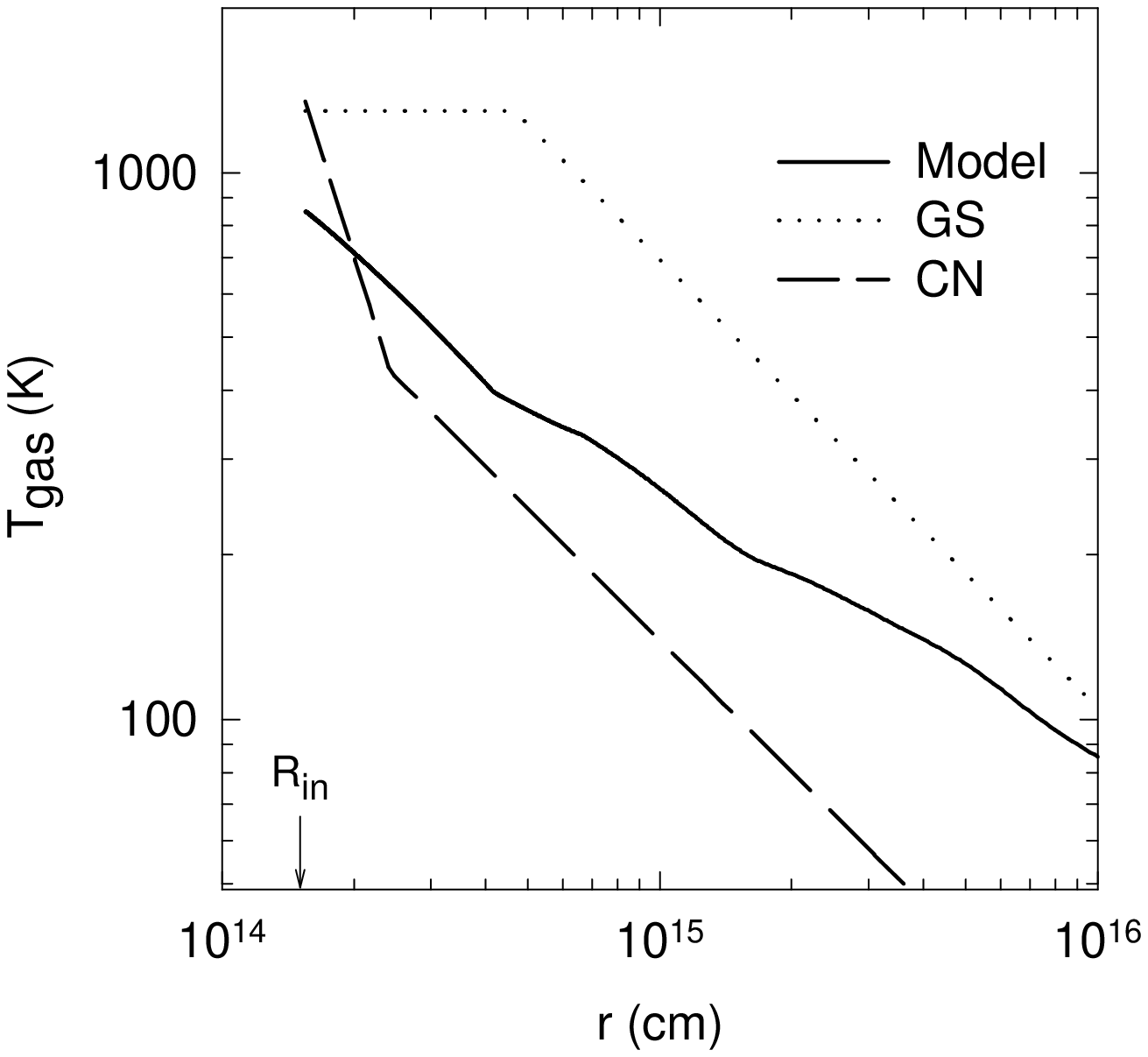}}
\figcaption[Tgas.ps] {The gas temperature profile from our best-fit model.
Shown are also power-law fits for the Goldreich \& Scoville (GS) and Chen \&
Neufeld (CN) temperature profiles.
     \label{fig:Tgas}}
\smallskip
}
%%%%%%%%%%%%%%%%%%%%%%%%%%%%%%%%%%%%%%%%%%%%%%%%%%%%%%%%%%%%%

The source of the large discrepancy with N96 is not clear. Neufeld et al
suggest that the temperature profile could be the reason, but this does not
seem to be the case. Our temperature profile is not that different from the CN
profile, which was used in the N96 study, especially in the relevant range $T
\ga 400$ K (figure \ref{fig:Tgas}). We compared the various contributions to
heating and cooling with those listed by CN and \citet{nk93}, and find good
agreement; the difference in resulting profiles can be attributed to the
different parameters used in the two calculations (CN used \Mdot\ = 3\x\E{-5}
\Mo\ yr$^{-1}$). The two profiles differ much more with the GS profile
employed in the B96 model, which produced \Mdot\ similar to ours. We suspect
that a more important source of difference could be the radiation field, since
radiative excitations play an important role in the water population
distribution. A comparison is impossible since N96 do not give details of the
radiation field they employed. A proper radiation field appears crucial for
the water line calculations.

The uncertainty in fitted parameters depends on the particular observations
that constrain them.  We estimate that the acceptable range for \rgd\ is
\about\ 850$\pm$100. We experimented with both power-law and single-size ($a$ =
0.1 \mic) grain distributions and found the differences negligible.  The large
uncertainties in the measured line fluxes translate into a large uncertainty in
the abundances of the two species of water.  Between the two, the para-H$_2$O
abundance is subject to the larger uncertainty and can vary by as much as
factor 3, so that the ortho:para ratio can be anywhere from 1 to 1/3. By
comparison, B96 obtained 1 for this ratio.  All of these results differ greatly
from the thermodynamic limit of 3. Depending on the ortho:para ratio, we find
acceptable models for $n({\rm H_2O})/n(\rm H_2)$ in the range \about\
1--4\x\E{-4}.

In the modeling efforts of both B96 and N96, \Mdot\ was one of numerous free
parameters fitted from the water line observations.  The scaling approach taken
here reduces the number of free parameters to the essential minimum and
determines \Mdot\ prior to the water line fitting.  Table 1 breaks the model
parameters into three categories.  Quantities in the first division are
specified as input, in addition to the IR and water observations. The second
group includes the free parameters of our two fitting procedures while the
third group lists results derived from our fits. Given grain properties, the IR
observations are fitted with the three free parameters \tV, \Tc\ and $Y$ and
the single independent input \Ts. The resulting model determines also the
dimensionless velocity profile. Adding as independent input the source distance
and the wind final velocity, the model results determine the full velocity
profile, and \tV\ determines also \Mdot\ and the gas-to-dust ratio. Fitting the
water lines involves no other input and only two free parameters --- the ortho-
and para-water abundances. Thanks to the central role of water lines in the gas
temperature calculation, the gas temperature is determined self-consistently as
part of this second fitting procedure.  Since the only free parameters in
fitting the observed water fluxes are the abundances of the two species,
confidence in the derived values is greatly enhanced. In principle, we could
have used \Mdot\ as a free parameter in the water calculations, as in the
previous studies.  In that case, consistency between the results of the two
fitting procedures would be used as an additional constraint, automatically met
by our calculation.

While radiation pressure on the dust grains is generally accepted as the
driving mechanism behind the wind expansion, this mechanism has not been fully
tested. The solutions of radiative transfer and the hydrodynamics problems
must result in the same parameters, but this fundamental test has not been
performed thus far; self-consistent modeling of both the IR emission and the
wind structure in the same source has not yet been attempted. The most
detailed previous calculation we are aware off is the TB modeling of R Cas.
However, in that work the radiation field was not calculated
self-consistently, instead it was derived from a dust temperature profile that
was assumed beforehand as an input property. In contrast, DUSTY determines
this temperature from a proper calculation of radiative equilibrium coupled to
the radiative transfer including dust scattering, absorption and emission. The
spectral energy distribution is fitted with just three free parameters, only
one of which (\tV) is significant. Once these parameters are set, the outflow
terminal velocity determines the entire velocity profile without any more
freedom in the model. It is highly significant that a single self-consistent
model with the minimal necessary number of parameters provides agreement with
both the spectral energy distribution and the molecular velocity observations.
Apart from resolving the \Mdot\ discrepancy in W Hya, the success of our model
provides strong support for the basic paradigm of winds in late-type stars.

\acknowledgements

We thank Michael Barlow for the ISO LWS spectrum of W Hya, David Neufeld for
help with the water molecular data, \v Zeljko Ivezi\'c for his comments on the
manuscript and the referee for useful suggestions and for pointing out the
Hipparcos distance measurement. The partial support of NASA and NSF is
gratefully acknowledged.

%%%%%%%%%%%%%%%%%%%%%%%%%%%%% Table 1 %%%%%%%%%%%%%%%%%%%%%%%%%%%%%
\let\n=\tablenotemark

\begin{table}[htbp]
\begin{center}
\caption{Model Parameters  \label{tab:tab1}} \centerline{}

\begin{tabular}{ll}
\tableline \tableline

\noalign{\medskip}
{\em Input parameters}\/:           &               \\
Stellar temperature\n{a}            & 2500 K        \\
Distance\n{b}                       & 115 pc        \\
Final velocity\n{c}                 & 8 km~s$^{-1}$ \\

\tableline \noalign{\medskip}
{\em Fitting parameters}\/:         &                \\
Optical depth at 0.55 \mic\n{I}     & 0.83           \\
Dust temperature at \Rin\n{I}       & 1000 K         \\
Shell thickness (\Rout/\Rin)\n{I}   & 12,000         \\
H$_2$O abundance at \Rin\n{w}       & 1.0\x\E{-4}    \\
ortho-H$_2$O:para-H$_2$O\n{w}       & 1:1.3          \\

\tableline \noalign{\medskip}
{\em Derived parameters}\/:         &                            \\
Stellar radius                      & 3.9\x\E{13} cm             \\
Stellar luminosity                  & 11,050 $L_{\odot}$         \\
Shell inner radius (\Rin)           & 1.6\x\E{14} cm             \\
Mass loss rate                      & 2.3\x\E{-6} \Mo\ yr$^{-1}$ \\
Gas-to-dust mass ratio              & 850                        \\
Shell outer radius (\Rout)          & 1.2 pc                     \\
Dust temperature at \Rout           & 17 K                       \\
\tableline
\\
\n{a}{ }\ \citet{hanif95}                      & \\
\n{b}{ }\ Hipparcos Catalog                    & \\
\n{c}{ }\ \citet{young95}                      & \\
\n{I}{ }\ fitted from IR observations          & \\
\n{w}{ }\ fitted from H$_2$O lines             & \\

\end{tabular}
\end{center}

\end{table}

%%%%%%%%%%%%%%%%%%%%%%%%%%%%%%%%%%%%%%%%%%%%%%%%%%%%%%%%%%%%%%%%%%%%%%%%%%%%%%

\begin{deluxetable}{rcrrc}
\tablewidth{0pt} \tablecaption{Observed and model fluxes of H$_2$O lines}
\tablehead{ \multicolumn{1}{c}{$\lambda$ (\mic)}
           & \colhead{Transition}
           & \multicolumn{1}{r}{$F_{\rm{mod}}$\n{a}}
           & \multicolumn{1}{r}{$F_{\rm{obs}}$\n{a}}
           & ${F_{\rm{mod}}\over F_{\rm{obs}}}$
}
\startdata
   180.486  & o: 2$_{ 2 1}$$\to$2$_{ 1 2}$  &   3.43 &   2.90 &    1.18 \\
   179.527  & o: 2$_{ 1 2}$$\to$1$_{ 0 1}$  &  12.48 &   8.66 &    1.44 \\
   174.624  & o: 3$_{ 0 3}$$\to$2$_{ 1 2}$  &   8.31 &   9.21 &    0.90 \\
   156.265  & o: 5$_{ 2 3}$$\to$4$_{ 3 2}$  &   2.42 &   5.41 &    0.45 \\
   144.517  & p: 4$_{ 1 3}$$\to$3$_{ 2 2}$  &   5.57 &   6.38 &    0.87 \\
   132.408  & o: 4$_{ 2 3}$$\to$4$_{ 1 4}$  &   2.97 &   5.50 &    0.54 \\
   125.356  & p: 4$_{ 0 4}$$\to$3$_{ 1 3}$  &  12.39 &  16.60 &    0.75 \\
   113.538  & o: 4$_{ 1 4}$$\to$3$_{ 0 3}$  &  11.27 &  17.40 &    0.65 \\
   108.073  & o: 2$_{ 2 1}$$\to$1$_{ 1 0}$  &  15.47 &  13.00 &    1.19 \\
    89.989  & p: 3$_{ 2 2}$$\to$2$_{ 1 1}$  &  18.11 &  27.30 &    0.66 \\
    83.283  & p: 6$_{ 0 6}$$\to$5$_{ 1 5}$  &  14.42 &  22.60 &    0.64 \\
    78.742  & o: 4$_{ 2 3}$$\to$3$_{ 1 2}$  &  14.08 &  28.00 &    0.50 \\
    67.089  & p: 3$_{ 3 1}$$\to$2$_{ 2 0}$  &  17.80 &  59.80 &    0.30 \\
    66.438  & o: 3$_{ 3 0}$$\to$2$_{ 2 1}$  &  18.45 &  22.90 &    0.81 \\
    63.458  & p: 8$_{ 0 8}$$\to$7$_{ 1 7}$  &  12.99 &  30.40 &    0.43 \\
    58.699  & o: 4$_{ 3 2}$$\to$3$_{ 2 1}$  &  15.35 &  19.50 &    0.79 \\
    57.637  & p: 4$_{ 2 2}$$\to$3$_{ 1 3}$  &  30.13 &  36.20 &    0.83 \\
    40.691  & o: 4$_{ 3 2}$$\to$3$_{ 0 3}$  &  36.07 &  23.00 &    1.57 \\
    31.772  & o: 4$_{ 4 1}$$\to$3$_{ 1 2}$  &  39.95 &  63.00 &    0.63 \\
    29.837  & o: 7$_{ 2 5}$$\to$6$_{ 1 6}$  &  30.72 &  32.00 &    0.96 \\
\noalign{\smallskip}
    37.984  & o: 4$_{ 4 1}$$\to$4$_{ 1 4}$  &  13.19 &  28.00 &    0.47\n{b}\\

\enddata
\tablecomments{Ortho transitions are marked by ``o'', para by ``p''. The
57--180 \mic\ lines are from \citet{barlow96}, the 29--41 \mic\ lines from
\citet{neufeld96}. $^a$Fluxes are in \E{-20} W cm$^{-2}$. $^b$This line is
severely contaminated by blending with two others.
}
       \label{tab:tab2}
\end{deluxetable}

\ifEmulate{\end{document}}\fi

\newpage
\section*{FIGURES}

\includegraphics[width=0.8\columnwidth]{spectrum.ps}

\figcaption[spectrum.ps] {The spectral energy distribution of W Hya. Data
indicated by filled squares (NIR) are from \citet{wilson72}; open squares
(IRAS) from \citet{hawkins90}; filled circles (JCMT) from \citet{veen95} and
\citet{walm91}. The thin solid line shows the ISO data of \citet{barlow96}.
The bold solid line is our model result.
      \label{fig:spectrum}}

\newpage

\includegraphics[width=0.8\columnwidth]{vgas.ps}

\figcaption[vgas.ps] {The model prediction for the gas velocity profile. Data
points (from Szymczak et al, 1998) show the H$_2$O and OH maser and CO thermal
emission from the wind.  The SiO maser emission originates in the extended
atmosphere.
   \label{fig:vgas}}

\newpage

\includegraphics[width=0.8\columnwidth]{Tgas.ps}

\figcaption[Tgas.ps] {The gas temperature profile from our best-fit model.
Shown are also power-law fits for the Goldreich \& Scoville (GS) and Chen \&
Neufeld (CN) temperature profiles.
     \label{fig:Tgas}}


\begin{thebibliography}{}
    \bibitem[Barlow et al.(1996)]{barlow96} Barlow, M.J., et al.
             1996, \aap, 315, L241 (B96)
    \bibitem[Chen \& Neufeld(1995)]{cn95} Chen, W., \& Neufeld, D.A. 1995,
             \apj, 453, L99 (CN)
    \bibitem[Elitzur(1992)]{elitzur92} Elitzur, M. 1992,
       Astronomical Masers, Dordrecht: Kluwer
    \bibitem[Goldreich \& Scoville(1976)]{gs76} Goldreich, P., \&
             Scoville, N. 1976, \apj, 205, 144 (GS)
    \bibitem[Green et al.(1993)]{green93} Green, S., Maluendes, S., \&
             McLean, A.D. 1993, \apjs, 85, 181
    \bibitem[Haniff et al.(1995)]{hanif95} Haniff, C.A., Scholz, M., \&
             Tuthill, P.G. 1995, \mnras, 276, 640
    \bibitem[Hawkins(1990)]{hawkins90} Hawkins, G.W. 1990, \aap, 229, L5
    \bibitem[Ivezi\'c \& Elitzur(1995)]{ie95} Ivezi\'c, \v Z., \&
             Elitzur, M. 1995, \apj, 445, 415
    \bibitem[Ivezi\'c \& Elitzur(1997)]{ie97} Ivezi\'c, \v Z., \&
             Elitzur, M. 1997, \mnras, 287, 799
    \bibitem[Ivezi\'c et al.(1999)]{dusty} Ivezi\'c, \v Z., Nenkova, M., \&
             Elitzur, M. 1999, User Manual for {\sc DUSTY},
             University of Kentucky Internal Report
    \bibitem[Laor \& Draine(1993)]{ld93} Laor, A., \& Draine, B.T. 1993,
             \apj, 402, 441
    \bibitem[Mathis et al.(1977)]{mrn} Mathis, J.S., Rumple, W., \&
             Nordsieck, K.H. 1977, \apj, 217, 425
    \bibitem[Neufeld \& Kaufman(1993)]{nk93} Neufeld, D.A., \& Kaufman, M.J.
             1993, \aap, 418, 263
    \bibitem[Neufeld et al.(1996)]{neufeld96} Neufeld, D.A., et al. 1996,
             \aap, 315, L237 (N96)
    \bibitem[Rothman et al.(1998)]{rothman98} Rothman, L.S., et al. 1998,
             JQSRT, 60, 665
    \bibitem[Szymczak et al.(1998)]{szymczak98} Szymczak, M., Cohen, R.J.,
             \& Richards, A.M.S. 1998, \mnras, 297, 1151
    \bibitem[Truong-Bach et al.(1999)]{truong99} Truong-Bach, et al. 1999,
             \aap, 345, 925 (TB)
    \bibitem[van der Veen et al.(1995)]{veen95} van der Veen, W.E.C.J.,
             Omont, A., Habing, H.J., \& Matthews, H.E. 1995, \aap, 295, 445
    \bibitem[Walmsley et al.(1991)]{walm91} Walmsley, C.M., Chini, R., 
Steppe, H.,
             Forveille, T., \& Omont, A. 1991, \aap, 248, 555
    \bibitem[Wilson et al.(1972)]{wilson72} Wilson, W.J., Schwatz,P.R.,
             Neugebauer, G., Harvey, P.M., \& Becklin, E.E. 1972,
             \apj, 177, 523
    \bibitem[Young(1995)]{young95} Young, K., 1995, \apj, 445, 872
\end{thebibliography}
\end{document}